\documentstyle[twoside,psfig]{article}

\catcode`\@=11
\long\def\@makefntext#1{
\protect\noindent \hbox to 3.2pt {\hskip-.9pt
$^{{\eightrm\@thefnmark}}$\hfil}#1\hfill}               

\def\thefootnote{\fnsymbol{footnote}}
\def\@makefnmark{\hbox to 0pt{$^{\@thefnmark}$\hss}}    

\def\ps@myheadings{\let\@mkboth\@gobbletwo
\def\@oddhead{\hbox{}
\rightmark\hfil\eightrm\thepage}
\def\@oddfoot{}\def\@evenhead{\eightrm\thepage\hfil
\leftmark\hbox{}}\def\@evenfoot{}
\def\sectionmark##1{}\def\subsectionmark##1{}}

\oddsidemargin=\evensidemargin
\renewcommand{\thefootnote}{\fnsymbol{footnote}}

\newcounter{sectionc}\newcounter{subsectionc}\newcounter{subsubsectionc}
\renewcommand{\section}[1] {\vspace{12pt}\addtocounter{sectionc}{1}
\setcounter{subsectionc}{0}\setcounter{subsubsectionc}{0}\noindent
        {\tenbf\thesectionc. #1}\par\vspace{5pt}}
\renewcommand{\subsection}[1] {\vspace{12pt}\addtocounter{subsectionc}{1}
        \setcounter{subsubsectionc}{0}\noindent
        {\bf\thesectionc.\thesubsectionc. {\kern1pt \bfit #1}}\par\vspace{5pt}}
\renewcommand{\subsubsection}[1] {\vspace{12pt}\addtocounter{subsubsectionc}{1}
        \noindent{\tenrm\thesectionc.\thesubsectionc.\thesubsubsectionc.
        {\kern1pt \tenit #1}}\par\vspace{5pt}}
\newcommand{\nonumsection}[1] {\vspace{12pt}\noindent{\tenbf #1}
        \par\vspace{5pt}}

\newcounter{appendixc}
\newcounter{subappendixc}[appendixc]
\newcounter{subsubappendixc}[subappendixc]
\renewcommand{\thesubappendixc}{\Alph{appendixc}.\arabic{subappendixc}}
\renewcommand{\thesubsubappendixc}
        {\Alph{appendixc}.\arabic{subappendixc}.\arabic{subsubappendixc}}

\renewcommand{\appendix}[1] {\vspace{12pt}
        \refstepcounter{appendixc}
        \setcounter{figure}{0}
        \setcounter{table}{0}
        \setcounter{lemma}{0}
        \setcounter{theorem}{0}
        \setcounter{corollary}{0}
        \setcounter{definition}{0}
        \setcounter{equation}{0}
        \renewcommand{\thefigure}{\Alph{appendixc}.\arabic{figure}}
        \renewcommand{\thetable}{\Alph{appendixc}.\arabic{table}}
        \renewcommand{\theappendixc}{\Alph{appendixc}}
        \renewcommand{\thelemma}{\Alph{appendixc}.\arabic{lemma}}
        \renewcommand{\thetheorem}{\Alph{appendixc}.\arabic{theorem}}
        \renewcommand{\thedefinition}{\Alph{appendixc}.\arabic{definition}}
        \renewcommand{\thecorollary}{\Alph{appendixc}.\arabic{corollary}}
        \renewcommand{\theequation}{\Alph{appendixc}.\arabic{equation}}
        \noindent{\tenbf Appendix \theappendixc #1}\par\vspace{5pt}}
\newcommand{\subappendix}[1] {\vspace{12pt}
        \refstepcounter{subappendixc}
        \noindent{\bf Appendix \thesubappendixc. {\kern1pt \bfit #1}}
        \par\vspace{5pt}}
\newcommand{\subsubappendix}[1] {\vspace{12pt}
        \refstepcounter{subsubappendixc}
        \noindent{\rm Appendix \thesubsubappendixc. {\kern1pt \tenit #1}}
        \par\vspace{5pt}}

\newcommand{\textlineskip}{\baselineskip=13pt}
\newcommand{\smalllineskip}{\baselineskip=10pt}

\def\eightcirc{
\begin{picture}(0,0)
\put(4.4,1.8){\circle{6.5}}
\end{picture}}
\def\eightcopyright{\eightcirc\kern2.7pt\hbox{\eightrm c}}

\newcommand{\publisher}[2]{{\begin{center}\footnotesize\smalllineskip
        Received #1\\
        Revised #2
        \end{center}
        }}

\def\abstracts#1#2#3{{
        \centering{\begin{minipage}{4.5in}\footnotesize\baselineskip=10pt
        \parindent=0pt #1\par
        \parindent=15pt #2\par
        \parindent=15pt #3
        \end{minipage}}\par}}

\newcommand{\bibit}{\nineit}
\newcommand{\bibbf}{\ninebf}
\renewenvironment{thebibliography}[1]
        {\frenchspacing
         \ninerm\baselineskip=11pt
         \begin{list}{\arabic{enumi}.}
        {\usecounter{enumi}\setlength{\parsep}{0pt}
         \setlength{\leftmargin 12.7pt}{\rightmargin 0pt} 
         \setlength{\itemsep}{0pt} \settowidth
        {\labelwidth}{#1.}\sloppy}}{\end{list}}

\newcounter{itemlistc}
\newcounter{romanlistc}
\newcounter{alphlistc}
\newcounter{arabiclistc}

\newcommand{\fcaption}[1]{
        \refstepcounter{figure}
        \setbox\@tempboxa = \hbox{\footnotesize Fig.~\thefigure. #1}
        \ifdim \wd\@tempboxa > 5in
           {\begin{center}
        \parbox{5in}{\footnotesize\smalllineskip Fig.~\thefigure. #1}
            \end{center}}
        \else
             {\begin{center}
             {\footnotesize Fig.~\thefigure. #1}
              \end{center}}
        \fi}

\newcommand{\tcaption}[1]{
        \refstepcounter{table}
        \setbox\@tempboxa = \hbox{\footnotesize Table~\thetable. #1}
        \ifdim \wd\@tempboxa > 5in
           {\begin{center}
        \parbox{5in}{\footnotesize\smalllineskip Table~\thetable. #1}
            \end{center}}
        \else
             {\begin{center}
             {\footnotesize Table~\thetable. #1}
              \end{center}}
        \fi}

\def\@citex[#1]#2{\if@filesw\immediate\write\@auxout
        {\string\citation{#2}}\fi
\def\@citea{}\@cite{\@for\@citeb:=#2\do
        {\@citea\def\@citea{,}\@ifundefined
        {b@\@citeb}{{\bf ?}\@warning
        {Citation `\@citeb' on page \thepage \space undefined}}
        {\csname b@\@citeb\endcsname}}}{#1}}

\newif\if@cghi
\def\cite{\@cghitrue\@ifnextchar [{\@tempswatrue
        \@citex}{\@tempswafalse\@citex[]}}
\def\citelow{\@cghifalse\@ifnextchar [{\@tempswatrue
        \@citex}{\@tempswafalse\@citex[]}}
\def\@cite#1#2{{$\null^{#1}$\if@tempswa\typeout
        {IJCGA warning: optional citation argument
        ignored: `#2'} \fi}}

\def\fnt#1#2{\footnotetext{\kern-.3em
        {$^{\mbox{\scriptsize #1}}$}{#2}}}

\def\fpage#1{\begingroup
\voffset=.3in
\thispagestyle{empty}\begin{table}[b]\centerline{\footnotesize #1}
        \end{table}\endgroup}

\def\runninghead#1#2{\pagestyle{myheadings}
\markboth{{\protect\footnotesize\it{\quad #1}}\hfill}
{\hfill{\protect\footnotesize\it{#2\quad}}}}
\font\tenrm=cmr10
\font\tenit=cmti10
\font\tenbf=cmbx10
\font\bfit=cmbxti10 at 10pt
\font\ninerm=cmr9
\font\nineit=cmti9
\font\ninebf=cmbx9
\font\eightrm=cmr8

\def\qed{\hbox{${\vcenter{\vbox{                        
   \hrule height 0.4pt\hbox{\vrule width 0.4pt height 6pt
   \kern5pt\vrule width 0.4pt}\hrule height 0.4pt}}}$}}

\renewcommand{\thefootnote}{\fnsymbol{footnote}}        

\begin{document}
\setlength{\textheight}{7.7truein}  

\runninghead{Quantum-Gravity Phenomenology: Status and Prospects}{
Quantum-Gravity Phenomenology: Status and Prospects}

\normalsize\textlineskip
\thispagestyle{empty}
\setcounter{page}{1}


\vspace*{0.88truein}

\fpage{1}
\centerline{\bf QUANTUM-GRAVITY PHENOMENOLOGY: STATUS AND PROSPECTS}
\vspace*{0.37truein}
\centerline{\footnotesize GIOVANNI AMELINO-CAMELIA}
\baselineskip=12pt
\centerline{\footnotesize\it Dipartimento di Fisica,
Universit\'{a} di Roma ``La Sapienza'', P.le Moro 2}
\baselineskip=10pt
\centerline{\footnotesize\it 00185 Roma, Italy}

\vspace*{0.225truein}

\publisher{(received date)}{(revised date)}

\vspace*{0.21truein}
\abstracts{Over the last few years part of
the quantum-gravity community has adopted a more optimistic
attitude toward the possibility of finding experimental contexts
providing insight on non-classical properties of spacetime.
I review those quantum-gravity phenomenology proposals which were
instrumental in bringing about this change of attitude,
and I discuss the prospects for the short-term future
of quantum-gravity phenomenology.
}{}{}



\vspace*{1pt}\textlineskip      
\section{Quantum Gravity Phenomenology}    
\vspace*{-0.5pt}
\noindent
The ``quantum-gravity problem" has been studied
for more than 70 years assuming that no guidance could
be obtained from experiments. This in turn led to
the assumption that the most promising
path toward the solution of the problem would be the
construction and analysis of very ambitious theories, some
would call them ``theories of everything", capable of solving at once
all of the issues raised by the coexistence of gravitation
(general relativity) and quantum mechanics.
In other research areas the abundant availability
of puzzling experimental data encourages theorists to
propose phenomenological models which solve the puzzles but are
conceptually unsatisfactory on many grounds.
Often those apparently unsatisfactory models
turn out to provide an important starting point
for the identification of the correct (and conceptually satisfactory)
theoretical description of the new phenomena.
But in this quantum-gravity research area, since there was
no experimental guidance, it was inevitable for theorists
to be tempted into trying to identify the correct theoretical
framework relying exclusively on some criteria
of conceptual compellingness.
Of course, tempting as it may seem, this strategy would not be
acceptable for a scientific endeavor. Even the most
compelling and conceptually satisfying theory could not
be adopted without experimental confirmation.
The mirage that one day within an ambitious
quantum-gravity theory one
might derive from first principles a falsifiable prediction for
the mundane realm of doable experiments gives some ``scientific
legitimacy" to
these research programmes, but this possiblity
never materialized, it may well be just a mirage.

\setcounter{footnote}{0}
\renewcommand{\thefootnote}{\alph{footnote}}

\subsection{Objectives of quantum gravity phenomenology}
\noindent
Over the last few years this author and a growing
number of research groups have attempted to tackle the
quantum-gravity problem with an approach which is more
consistent with the traditional strategy of scientific work.
Simple (in some cases even simple-minded)
non-classical pictures of spacetime are being analyzed
with strong emphasis on their observable predictions.
Certain classes of experiments have been shown to
have extremely high sensitivity to
some non-classical features of spacetime.
We now even have (see later)
some first examples of experimental puzzles
whose solution is being sought also within simple ideas involving
non-classical pictures of spacetime.
The hope is that by trial and error, both on the theory side
and on the experiment side, we might eventually stumble upon
the first few definite (experimental!) hints on the quantum-gravity
problem.
Here I intend to give an overview
of this ``quantum gravity phenomenology"\cite{polonpap}.

``Quantum gravity phenomenology"
is an intentionally~\cite{polonpap} vague name,
reflecting the
fact that this new approach to quantum-gravity research
requires a combination of theory and experiments
and also reflecting the fact that it does not adopt any
particular prejudice concerning the structure of spacetime
at short distances (in particular, ``string
theory"~\cite{string1,string2}, ``loop quantum
gravity"~\cite{crLIVING,leebook,ashtNEW}
and ``noncommutative geometry"~\cite{connesbook,majidbok}
are seen as equally deserving
mathematical-physics programmes).
It is rather the proposal that quantum-gravity research should
proceed just in the old-fashioned way
of scientific endeavors: through small incremental
steps starting from what we know and combining mathematical-physics
studies with experimental studies to reach deeper and deeper layers
of understanding of the problem at hand (in this case
the short-distance structure of spacetime and the laws that govern it).

The most popular quantum-gravity approaches, such as
string theory and loop quantum gravity, could be described as ``top-to-bottom
approaches" since they start off with some key assumption about
the structure of spacetime at scales that
are some 17 orders of magnitude
beyond the scales presently accessible experimentally, and then they should
work their way back to the realm of doable experiments.
With ``quantum gravity phenomenology" I would like to refer to all
studies that are intended to contribute
to a ``bottom-to-top approach"
to the quantum-gravity problem.

Since the problem at hand is really difficult (arguably the most challenging
problem ever faced by the physics community)
it appears likely that the two complementary
approaches might combine in a useful way: for the ``bottom-to-top approach"
it is important to get some guidance from the (however tentative)
indications emerging
from the ``top-to-bottom approaches", while for ``top-to-bottom approaches"
it might be useful to be alerted by quantum-gravity phenomenologists
with respect to the type of new effects that could be most stringently
tested experimentally (it is hard for ``top-to-bottom approaches" to
obtain a complete description of ``real" physics, but perhaps it
would be possible to dig out predictions on some specific spacetime features
that appear to deserve special attention in light of the corresponding
experimental sensitivities).

Until very recently the idea of a quantum-gravity phenomenology,
and in particular of attempts of identification of experiments with
promising  sensitivity, was very far from the interests of
mainstream quantum-gravity researchers.
This is still true for a significant portion of the community,
but finally, just over the last couple of years,
there is also a significant portion of
the community which is forming an interest in
experiment-aimed research


\subsection{Prehistory of quantum gravity phenomenology}
\noindent
To this author's knowledge the first experiment-related studies
with some relevance for the quantum-gravity problems
are the ones pertaining the effects of {\bf classical}
gravitational fields on matter-interferometry experiments.
This really started in the mid 1970s with the renowned
experiment performed by Colella, Overhauser and Werner~\cite{cow}.
That experiment has been followed by several modifications
and refinements (often labeled ``COW experiments'' from the initials
of the scientists involved in the first experiment)
all probing the same basic physics, {\it i.e.} the validity
of the Schr\"{o}dinger equation
\begin{eqnarray}
\left[ - \left( {\hbar^2 \over 2 \, M_I} \right) \vec{\nabla}^2
+ M_G \, \phi(\vec{r}) \right] \psi(t,\vec{r}) = i \, \hbar \,
{\partial \, \psi(t,\vec{r}) \over \partial t}
\label{coweq}
\end{eqnarray}
for the description of the dynamics of matter (with wave
function $\psi(t,\vec{r})$) in presence of the Earth's
gravitational potential $\phi(\vec{r})$.
[In (\ref{coweq}) $M_I$ and $M_G$ denote the inertial
and gravitational mass respectively.]

The COW experiments exploit the fact that
the Earth's gravitational potential puts together the contributions
of a very large number of particles (all the
particles composing the Earth)
and as a result, in spite of its per-particle weakness,
the overall gravitational field is large enough to introduce observable
effects.\footnote{Actually the effect turns out to be observably large
because of a double ``amplification": the first, and most significant,
amplification is the mentioned coherent addition of gravitational
fields generated by the particles that compose the Earth,
the second amplification~\cite{dharamCOW1} involves the ratio between
the wavelength of the particles used in the COW experiments
and some larger length scales involved in the experimental
setup.}.
This type of experiment of course does not probe any non-classical
property of spacetime. It is the classical gravitational field
that playes a role in the experiment.
In this sense it should be seen as a marginal aspect of quantum-gravity
phenomenology, just testing the correctness of (inherently robust)
ideas on the behaviour of quantum mechanics in curved (but still
classical) spacetime\footnote{While ``quantum-gravity phenomenology"
is being adopted to describe experiments aimed at detecting
quantum properties of spacetime, the wider subject of the interplay
between quantum mechanics and general relativity (even when spacetime
can be described in fully classical manner) is sometimes
called~\cite{iqgr} ``Interface of Quantum and Gravitational Realms"}.
However, some important insight for quantum-gravity research
has been gained through these experiments, particularly
with respect to the faith of the Equivalence Principle.
This subject deserves a dedicated review by the experts.
I here just bring to the reader's attention some
useful reading material~\cite{sakurai,gasperiniEP,dharamEP},
a recent experiment which appears to indicate a violation
of the Equivalence Principle~\cite{cowEPviol}
(but the reliability of this experimental result is still
being debated),
and some ideas for intruiging
new experiments~\cite{dharamCOW1,dharamCOW2}
of the COW type.

In the mid 1980s, together with the analysis of other experimental
contexts~\cite{anan1,anan2}
probing the interplay
between classical general relativity and quantum mechanics of
nongravitational degrees of freedom,
the analysis of a second class
of experimental contexts relevant for quantum gravity
was started. This second class of experiments is based on the
realization~\cite{ehns,huetpesk,kostcpt,emln,floreacpt}
that the sensitivity of CPT-symmetry tests using
the neutral-kaon and neutral-B systems\footnote{In addition to
the neutral-kaon and neutral-B systems
there has been recent discussion~\cite{ahlucpt,muracpt}
of the possibility to use neutrino physics in the study of
quantum-gravity-induced CPT violation.}$~$
is reaching a level such that even small quantum-gravity-induced
CPT violation might in principle be revealed.
Until now there is no evidence of any violation of the CPT symmetry.
It should also be noticed that the theory work
in this research line was not in the spirit
here advocated for quantum-gravity phenomenology.
In fact,
the CPT tests were motivated with one or another specific
idea about an ambitious (top-to-bottom) theory:
a certain version of noncritical string theory in the case
of Refs.~\cite{ehns,emln} and a certain perspective on
string field theory in the case of Ref.~\cite{kostcpt}.
Rather than a study of deviations from CPT symmetry
with the general objective of reflecting the variety of scenarios
by which this could come about in quantum gravity,
the motivation provided to the theory community and
to experimentalists was linked directly to a specific
theory-of-everything picture.
Moreover, the phenomenology that got set up in this
research line made direct reference only to particle physics,
and it was unclear which type of ideas about the {\bf structure}
of spacetime were being investigated (one thing is to search
for the CPT implications of, say, some schemes for spacetime
discreteness or noncommutativity, which would appeal to all those
involved in research in related schemes, another is to confront
the quantum-gravity community with some particle-physics
phenomenology whose connection to gravity/spacetime quantization
is not at all transparent or direct).
These aspects of those early CPT studies might have played a role
in the fact that this research line did not manage to
affect the attitude toward experimental tests
of the quantum-gravity community.

\subsection{The dawn of quantum gravity phenomenology}
\noindent
The early pioneering development
of the research lines mentioned in the previous subsection
(research lines which are still to be deemed
crucial for quantum-gravity research)
did not lead to a change of attitude
of the quantum-gravity community. This change of attitude is
however materializing in these last few years, as indicated, for example,
by the sharp change of emphasis that one finds in comparing
authoritative quantum-gravity reviews published up to the
mid 1990s (see, {\it e.g.}, Ref.~\cite{chrisreview})
and the corresponding reviews published over the last couple of
years~\cite{leebook,ashtereview,crHISTO,leePW,gacqm100,carlip}.
Over the last few years several new ideas for tests
of quantum-gravity physics have appeared at increasingly fast pace,
with a fast growing (although, of course, still relatively small)
number of research groups joining the quantum-gravity-phenomenology
endeavor.
Of course the emergence of some first examples
(see later)
of experimental puzzles
whose solution can plausibly be sought within quantum-gravity
phenomenology marked an important turning point.

We now have several examples of experimentally accessible
contexts in which conjectured quantum-gravity effects are being
considered, including studies of in-vacuo dispersion using gamma-ray
astrophysics~\cite{grbgac,billetal},
studies of laser-interferometric limits on quantum-gravity induced
distance fluctuations~\cite{gacgwi,bignapap,nggwi,bignapatwo},
studies of the role of quantum-gravity effects in the determination
of the energy-momentum-conservation threshold conditions
for certain particle-physics processes~\cite{kifu,aus,gactp,jaco},
and studies of the role of quantum gravity in the determination
of particle-decay amplitudes~\cite{gacpion}.
These experimental contexts (together with the CPT tests
mentioned in the previous subsection) could be seen as the cornerstones
of quantum-gravity phenomenology since they are as close as one can
get to direct tests of space-time properties, such as space-time
symmetries. I postpone to Sections~3 and 4 a discussion
of these ideas.

In closing this subsection I should mention (even though I shall
not come back to these studies in the rest of the paper)
that there are also other experimental proposals that are
part of the quantum-gravity-phenomenology programme but rely
on the mediation of some dynamical theory
of matter in quantum space-time,
so that negative results of these experimental searches might
not teach us much at the qualitative level about spacetime
structure (the predicted magnitude of the effects
depends on spacetime features just as much as it depends
on some features of the postulated description of the dynamics
of matter in that spacetime).
Interested readers can find descriptions of these proposals
in Refs.~\cite{polonpap,veneziano,peri,garaytest,lamer}.

\subsection{Identification of experiments}
\noindent
The first step for the identification of experiments
relevant for quantum gravity is of course the identification
of the characteristic scale of this new physics.
This is a point on which we have relatively robust
guidance from theories and theoretical arguments:
the characteristic scale at which non-classical properties
of spacetime physics become large (as large as the classical properties
they compete with) should be\footnote{I do not
review here the arguments that single out the Planck length.
There is a large number of, apparently indepedent,
arguments that all converge to this scale. Consistently with
the overall attitude adopted
in quantum-gravity phenomenology one should maintain some
level of healthy doubt also about these arguments,
but among all theory indications it is certainly fair to
say that the indication of this characteristic scale
is the most robust element of guidance for quantum-gravity phenomenology.
Still one should take notice of recent studies~\cite{led}
finding ways
to effectively increase the size of this characteristic length scale.
Those arguments are not in any way ``natural" (not even in the eyes of
the scientists who proposed them) but they do justify some
reason of concern that perhaps we cannot even rule out
surprises about the characteristic scale of quantum-gravity effects.}$~$
the Planck length $L_p \sim 10^{-35}m$ (or equivalently
its inverse, the Planck scale $E_p \sim 10^{28} eV$).

The next step is the identification of the
type of effects that quantum-gravity theories
might predict. Unfortunately, in spite of more than 70 years of theory
work on the quantum-gravity problem, and a certain proliferation
of theoretical frameworks being considered, there is only
a small number of physical effects that have been considered
within quantum-gravity theories. Moreover, most of these effects
concern strong-gravity contexts, such as black-hole physics
and big-bang physics, which are exciting at the level
of conceptual analysis and development of formalism,
but of course are not very promising
for the actual (experimental) discovery of manifestations of
non-classical properties of spacetime.
For example,
the fact
that we are not even able to observe/verify the expected
classical properties of black holes clearly suggests that
this is not a promising context
for quantum-gravity phenomenology\footnote{But it does make sense to use
related observational facts to constrain
quantum-gravity theories: for example, some theories might
be rejected if found to be inconsistent with what we,
in some sense, ``know" about
the early universe or with future data on
the abundance of black holes.}.

We clearly should give priority to quantum-gravity effects that
modify our description of {\bf flat} spacetime. The effects will perhaps
be less significant than, say, in black hole physics (in some
aspects of black hole physics quantum-gravity effects
might be as large as classical physics effects), but we are
likely to be better off considering flat spacetime,
in which the quality of the data we can obtain is extremely high,
even though this will cost us a large suppression of
quantum-gravity effects,
a suppression which is
likely to take the form of some power of the ratio
between the Planck length and the wavelength of the
particles involved.

The presence of these suppression factors
on the one hand reduces sharply our chances of finding
quantum-gravity effects, but on the other hand simplifies
the problem of identifying promising experimental contexts,
since these experimental contexts must enjoy very
special properties which would not go easily unnoticed.
For laboratory experiments
even an optimistic estimate of these suppression factors
leads to a suppression of order $10^{-16}$, which one obtains
by assuming (probably already using some optimism)
that at least some quantum-gravity effects are only linearly
suppressed by the Planck length and taking as particle wavelength
the shorter wavelengths we are able to produce ($\sim 10^{-19}m$).
In astrophysics (which however limits one to ``observations"
rather than ``experiments") particles of shorter wavelength
are being studied, but even for the highest energy cosmic rays,
with energy of $\sim 10^{20}eV$ and therefore wavelengths
of $\sim 10^{-27}m$, a suppression of the type $L_p/\lambda$
would take values of order $10^{-8}$.
It is mostly as a result of this type of considerations
that traditional quantum-gravity reviews considered
the possibility of experimental studies with unmitigated
pessimism~\cite{chrisreview}.
However, the presence of these large suppression factors
surely cannot suffice for drawing any conclusions.
Even just looking within the subject of particle physics
we know that certain types of small effects can be studied,
as illustrated by the example of the remarkable limits
obtained on proton instability. Outside of fundamental physics more
success stories of this type are easily found: think for example
of brownian motion.
It is hard but clearly not impossible to find
experimental contexts in which there is effectively an
amplification of the small effect one intends to study.
The prediction of proton decay within certain grandunified theories
of particle physics is really a small effect, suppressed by the fourth
power of the ratio between the mass of the proton and
grandunification scale, which is only
three orders of magnitude smaller than the Planck scale.
In spite of this horrifying suppression,
of order $[m_{proton}/E_{gut}]^4 \sim 10^{-64}$,
with a simple idea we have managed to acquire full sensitivity
to the new effect: the proton lifetime predicted by grandunified
theories is of order $10^{39}s$ and ``quite a few" generations
of physicists should invest their own lifetimes staring at a single
proton before its decay, but by managing to keep under observation
a large number of protons (think for example of a situation in which
$10^{33}$ protons are monitored)
our sensitivity to proton decay is dramatically increased.
In that context the number of protons is the (ordinary-physics)
dimensionless quantity that works as ``amplifier" of the new-physics
effect. Similar considerations explain the success of brownian-motion
studies already a century ago.

We should therefore focus our attention\cite{polonpap}
on experiments which have something to do with spacetime
structure (in flat-spacetime situations only structure
can be revealed, in the sense discussed in the following sections)
and that host an ordinary-physics dimensionless quantity
large enough that (if we are ``lucky") it could amplify
the extremely small effects we are hoping to discover.
So there is clearly a first level of analysis
in which one identifies experiments with this rare
quality, and a second level of analysis in which one
tries to establish whether indeed the candidate ``amplifier"
could possibly amplify effects connected with spacetime
structure.

In parallel with this type of analysis one can perform
a complementary analysis which considers one-by-one
the quantum-gravity effects which (for good or bad reasons)
have surfaced in the quantum-gravity literature,
and then for each of them attempts to identify
the most significant experimental limit which can be obtained
with available technologies.
Only very few quantum-gravity effects that might affect
flat-spacetime physics (what we presently perceive as physics
occurring in the structureless arena of flat Minkowski spacetime)
have surfaced in the literature. Moreover, for some of these effects
one can quickly realize that the likelyhood of finding
an ``amplifier" is vanishingly small. For example, many studies
consider discretizations of the concepts of length, area, volume,
but all of these pictures predict geometric quanta whose magnitude
is set by the Planck length independently of the size of the
geometric observable being analzed (no amplification in going
from the study of, say, small areas to the study of large areas).
However, the idea of discretization, with associated short-distance
nonlocality, does encourage the idea
of departures from conventional CPT symmetry, since the
CPT theorem relies on absolute locality (locality at all scales).
Similarly spacetime noncommutativity, another justifiably popular
quantum-gravity idea, also encourages CPT studies
because it is quite natural~\cite{gacmajid} (though not necessary)
to find that P and/or T transformations acquire
new properties in a given noncommutative spacetime.
CPT-symmetry studies have the disadvantage that
in some approaches to the quantum-gravity problem, {\it e.g.}
loop quantum gravity, one is not yet able to couple ordinary
particles to gravity, and the theories are therefore
unprepared to describe $C$ transformations.

Planck-scale discreteness or noncommutativity
also provide encouragement for tests of Lorentz symmetry.
The continuous symmetries of a spacetime reflect of course
the structure of that spacetime. Ordinary Lorentz symmetry
is governed by the single scale that sets the structure
of classical Minkowski spacetime, the speed-of-light
constant $c$. If one introduces additional structure
in a flat spacetime its symmetries will be accordingly
affected. This is particularly clear in certain classes
of noncommutative spacetimes, whose symmetry transformations
are characterized by the noncommutativity length scale\cite{gacdsr}
(possibly the Planck scale) in addition to $c$, and infinitesimal
symmetry transformations are actually described in terms
of the new language of Hopf algebras\cite{majrue,kpoinap},
rather than by the Poincar\'{e} Lie algebra.
Because of their sensitivity to any type of structure
introduced in the description of flat spacetime,
tests of the continuous Lorentz symmetry will probably be found to
be relevant for the majority of quantum-gravity approaches.

In addition to discreteness and noncommutativity, with associated
possible deviations from conventional Lorentz and CPT symmetry,
another picture of spacetime which would have significant
implications for our description of flat spacetime
is the one of ``spacetime foam", which
has frequently surfaced
in the quantum-gravity literature,
although always described at a sort of intuitive formal level,
without proper operative description of any associated effects.
Since experiments can
only test well defined predictions,
the fact that
the description of spacetime foam that one finds in traditional
quantum-gravity studies is not operative of course confronts
quantum-gravity phenomenology with a first obvious task
of providing such an operative definition.
This operative definition must capture the indications
that come from formalism, {\it i.e.} must give physical
characterizations of the vague concept of ``fuzzy geometry" which
is the characterizing property advocated in studies of
spacetime foam.
In spacetime foam
certain sharp predictions of classical-spacetime physics
are rendered ``unsharp" by quantum-gravity effects.
As a way to capture (and endow with physical reality)
the concept of fuzzy distance,
this author introduced in Refs.\cite{gacgwi,bignapap,bignapatwo}
a first operatively defined characteristic of spacetime foam:
the noise levels in the readout of a laser interferometer
would receive an irreducible (fundamental) contribution
from quantum-gravity effects.
This noise can in principle be reduced to zero in classical physics,
while
the ordinary quantum properties of matter already introduce an
extra noise contribution with respect to classical physics.
Spacetime foam would introduce another source
of noise, reflecting the fact that the distances involved in
the experiment would be inherently unsharp in a
foamy spacetime picture.

In general this idea of spacetime foam motivates us to seek
effects that modify classical-spacetime physics
in a nonsystematic way.
The studies of Lorentz symmetry and CPT symmetry mentioned above
motivate the search of systematic departures
from classical-spacetime physics; for example, in certain noncommutative
geometries the relevant concept of Lorentz symmetry introduces
a systematic dependence of the speed of photons on their wavelength.
It appears meaningful to introduce a terminology that would instead
attribute to ``spacetime foam"
all nonsystematic quantum-gravity effects.
(Actually this could be a good physical definition of spacetime foam.)

All the, now numerous, proposals that presently compose
quantum-gravity phenomenology can be recognized as belonging
to one of these three categories:
CPT-symmetry tests, Lorentz-symmetry tests, and searches
of foam-induced effects. This reflects the fact that
these are the only possibilities that have been discussed in
the quantum-gravity literature as candidate effects that could
be present in the spacetimes that we presently perceive as flat
and classical (Minkowski).
This may change in the future as more theory ideas are explored
and more experimental studies are considered.
But in general it will always be possible to
distinguish between tests/studies of systematic quantum-gravity
effects and nonsystematic quantum-gravity effects.
Systematic quantum-gravity
effects and nonsystematic quantum-gravity effects
have been introduced
in this Subsection at an intuitive level, but
they will be more
technically characterized in the next three Sections.

\subsection{Mathematical-physics aspects of
quantum-gravity phenomenology}
\noindent
One peculiarity of quantum-gravity phenomenology
with respect to other phenomenological programmes
(think for example of particle-physics phenomenology)
is that even some of the relatively unambitious nonclassical
pictures of spacetime and gravity that one should consider
in quantum-gravity phenomenology
may require a rather sophisticated level of mathematical analysis.

A good prototype example of theory work in quantum-gravity
phenomenology is provided by the study of flat noncommutative
spacetimes.
Clearly the introduction of a flat noncommutative spacetime
could not possibly provide a complete solution to the
quantum-gravity problem, but some of the top-to-bottom approaches
appear to indicate that the emergence of noncommutative
geometry in quantum gravity is plausible, and if spacetime
geometry is in general noncommutative then in particular
the spacetimes we presently perceive as classical continuous
and commutative (Minkowski spacetime) should then be described at
the fundamental level in terms of noncommutative geometry.
It is therefore legitimate for a bottom-to-top
approach to the quantum-gravity problem to consider (together
with other possibilities of course) the possibility
of noncommutative versions of Minkowski spacetime.
Starting ``from the bottom" it is difficult to favour
one version of noncommutative Minkowski over another,
but one can attack the problem in stages, starting
with the simplest cases. In fact, most work on noncommuative
versions of Minkowski spacetime has been on the two simplest
possibilities: canonical spacetimes, in which the commutators
of the spacetime coordinates are coordinate independent
\begin{equation}
\left[x_\mu,x_\nu\right] = i \theta_{\mu \nu}
\label{canodef}
\end{equation}
($\mu,\nu,\beta = 0,1,2,3$),
and Lie-algebra noncommutative spacetimes,
in which the commutators
of the spacetime coordinates are linear in the coordinates
\begin{equation}
\left[x_\mu,x_\nu\right] = i C^\beta_{\mu \nu} x_\beta ~.
\label{liedef}
\end{equation}
Among Lie-algebra noncommutative versions of Minkowski spacetime the
desire to preserve $O(3)$ space-rotation covariance
has focused most work on $\kappa$-Minkowski spacetime,
with a single deformation scale $\kappa$
\begin{equation}
\left[x_m,t\right] = {i \over \kappa} x_m ~,~~~~\left[x_m, x_l\right] = 0
\label{kmindef}
\end{equation}
($l,m = 1,2,3$).

We are finding out that in order to establish what are the characteristic
physical predictions of these spacetime pictures
some severe mathematical challenges must be faced.
For example, it has been realized that
in canonical noncommutative spacetimes
the Wilson decoupling
between high-energy and low-energy physics
does not hold\cite{seibIRUV,sussIRUV},
and the technical and conceptual understanding of the implications
of this delicate mathematical property is still
in progress\cite{dineIRUV,gacluisa}.
Motivation for these studies is coming also from a top-to-bottom
approach, since canonical noncommutative spacetimes
are used in effective-theory descriptions\cite{seibIRUV,sussIRUV}
of some features of the physics of strings in certain backgrounds.

Another example of delicate mathematical analysis needed
for extracting the physical predictions of these noncommutative spacetimes
is the study of the symmetries of $\kappa$-Minkowski.
It was realized that infinitesimal symmetry transformations
in this spacetime could be described in terms
of a Hopf algebra~\cite{gacmajid,kpoinap},
but for a few years it appeared that several\cite{kpoinap} Hopf algebras
could describe these symmetries, and, even more concerning,
it appeared that the (appropriately deformed)
infinitesimal Lorentz transformations
could not be combined to obtained a symmetry group of
finite deformed-Lorentz transformations.
This impasse was overcome only recently by identifying the
right Hopf algebra and finding that the associated
finite symmetry transformations do form group\cite{gacdsr}.
The possibility of these deformed symmetries can be tested
with forthcoming experiments (see Section~3).

The example of flat noncommutative spacetimes
is representative of other bottom-to-top
mathematical-physics studies aimed at experiments,
in the spirit of quantum-gravity phenomenology.
Another representative example is the study of the physics
of ``weave states of spacetime geometry".
These geometry descriptions emerged as part of the loop quantum
gravity research programme, but they were taken as bottom-to-top
starting point in various studies aimed at experiments
(see, {\it e.g.}, Refs.~\cite{gampul,mexweave}).

\section{Characterization of Systematic and Nonsystematic Quantum-Gravity
Effects}
\noindent
Systematic quantum-gravity
effects and nonsystematic quantum-gravity effects
have been introduced
in the previous Section at an intuitive level.
It is useful to characterize these concepts more
quantitatively with the help of a specific example.

Of course we want to focus on a crisp spacetime feature.
Let us consider the propagation of massless particles over a
distance $L$ in flat spacetime. In classical physics the distance $L$
would be classical, the massless particles would be point-like
and follow the classical trajectory along $L$.
In classical physics a
bunch of such particles, with energies however different among them,
which were emitted along the $x$ axis
simultaneously at time $t=0$ from position $(x_0,0,0)$
would reach simultaneously at time $t=L/c \equiv T$
the position $(x_0+L,0,0)$.

This is a good setup because it involves propagation through spacetime,
which could pick up features of spacetime structure,
and it involves Lorentz symmetry
through the speed-of-light constant $c$ and the wavelength-independence
of the time of travel $T$.

Ordinary (known) quantum properties of matter (in classical spacetime)
already modify this picture: quantum-mechanical uncertainties
impose that the time of emission of a particle of energy $E$ can
only be controlled with accuracy $1/E$, and there is of course
a corresponding limitation on how accurately the simultaneity of
the times of arrival can be established, but the relation $T=L/c$
will emerge if appropriate averaging over a large number of
observations is performed.

In this setup,
while classical physics (of particles and spacetime)
predicts the relation $T=L/c$ (a systematic relation
between the observables $T$ and $L$),
the quantum properties of the particles
(still assuming classicality of the spacetime)
introduce a nonsystematic effect, an uncertainty: $T = L/c {\pm} \delta T_{QM}$.
How could quantum gravity affect this prediction?
In order to be covered on all possible fronts we should be open
to the possibility of both systematic and nonsystematic quantum-gravity
effects. This can be captured in the formula
\begin{equation}
T = (L/c + \Delta T_{QG}) {\pm} \delta T_{QM} {\pm} \delta T_{QG} ~,
\label{generalTL}
\end{equation}
with self-explanatory notation.

At low energies we have very good access to this type of observations
and the associated large statistics allows us to draw
the safe conclusion that
\begin{equation}
lim_{E \rightarrow 0} \Delta T_{QG} =0 ~,
\label{limlowe}
\end{equation}
but the data available to us do not allow us to exclude
that $\Delta T_{QG} \neq 0$ at high energies.
This would of course require a deviation from ordinary
Lorentz invariance of the spacetime, a systematic effect
(an effect on the observed average value of $T$).
It turns out that such a systematic effect is predicted
by certain non-classical pictures of spacetime.
For example, according to the deformed Lorentz symmetries
of $\kappa$-Minkowski spacetime
one would predict\cite{gacdsr} (for $\kappa = 1/L_p$)
$\Delta T_{QG} \simeq L_p E T$ at energies $E$ small compared to $1/L_p$.

Of course, our low-energy observations also constrain
the admissable values of $\delta T_{QG}$ at low energies,
and in particular we can safely assume $\delta T_{QG} < \delta T_{QM}$
at low energies. Because of the nature of uncertainties it
would always be extremely hard to find evidence of the
contribution $\delta T_{QG}$ in physical contexts
such that $\delta T_{QG} < \delta T_{QM}$, but we might
eventually discover a physical context in which
the ``fuzziness of spacetime" dominates over the
ordinary uncertainties
of quantum mechanics: $\delta T_{QG} > \delta T_{QM}$.
This would allow us to establish a fuzzy feature of spacetime
itself. I propose to refer to all such nonsystematic
quantum-gravity effects as ``spacetime foam effects",
since their nature is consistent with the intuition
emerging from formal work on foam.

The issues I just introduced within
the example of the relation between $T$ and $L$ are of
course present in the analysis of all relations between
spacetime-related observables.
In another case the observables $A,B,C$ will be classically
related by $A= f_{classic}(B,C)$.
Ordinary quantum mechanics will often introduce (unless
the observables all commute with each other)
an uncertainty:
$A= f_{classic}(B,C) + \delta A_{QM}$.
Systematic quantum-gravity effects, in particular
deviations from classical symmetries, may modify
the bare relation between the observables,
$f_{classic}(B,C) \rightarrow f_{QG}(B,C;L_p)
= f_{classic}(B,C) + \Delta A_{QG}$,
and in addition nonsystematic (fuzzyness, foaminess) quantum-gravity
effects may introduce an additional source of uncertainty:
$A= (f_{classic}(B,C)+ \Delta A_{QG}) {\pm} \delta A_{QM} {\pm} \delta A_{QG}$.

\section{Lorentz Tests: an example of Studies of systematic Quantum-Gravity
Effects}
\noindent
If the Planck length, $L_p$, only has the role we presently attribute
to it, which is basically the role of a coupling constant
(an appropriately rescaled version of the coupling $G$),
no problem arises for FitzGerald-Lorentz contraction,
but if we try to promote $L_p$ to the status of an intrinsic
characteristic of space-time structure (or a characteristic of
the kinematic rules that govern particle propagation in space-time)
it is natural to find conflicts with FitzGerald-Lorentz contraction.

For example, it is very hard (perhaps even impossible)
to construct discretized versions or non-commutative versions
of Minkowski space-time which enjoy ordinary
Lorentz symmetry.
Pedagogical illustrative examples of
this observation have been discussed, {\it e.g.},
in Ref.~\cite{hooftlorentz} for the case of discretization
and in Refs.~\cite{majrue,kpoinap}
for the case of non-commutativity.
Under ordinary Lorentz boosts,
discretization length scales and/or non-commutativity length
scales naturally end up acquiring different values for
different inertial observers, just as one would expect
in light of the mechanism of FitzGerald-Lorentz contraction.
Recently there has been strong interest in the deviations
from ordinary Lorentz symmetry that emerge
in canonical noncommutative spacetimes\cite{sussIRUV,dineIRUV,gacluisa}
and in $\kappa$-Minkowski noncommutative spacetime\cite{gacdsr}.
As mentioned, the example of canonical noncommutative spacetimes is
also indirectly relevant for string theory.
Within loop quantum gravity deviations from ordinary Lorentz invariance
have been considered in particular in Refs.\cite{gampul,mexweave}.

Some dynamical mechanisms (of the spontaneous symmetry-breaking
type) that can lead to deviations from ordinary Lorentz invariance
have been considered in string
field theory~\cite{kosteLORENTZ} and in certain noncritical string-theory
scenarios\cite{aemn}.

Even outside mainstream quantum-gravity approaches
interest in Planck-scale deviations from Lorentz invariance is
growing
(see, {\it e.g.}, Refs.~\cite{thooftLIV,laughLIV}).

In this Section I want to show that even very small,
Planck-length suppressed, deviations from Lorentz invariance
could be within the reach of ongoing and forthcoming experiments.
Let us focus on the possible emergence of deformed dispersion relations
(which is present in the large majority of quantum-gravity-motivated
schemes for deviations from ordinary Lorentz invariance)
and let us just consider
the possibility that the standard dispersion
relation $E^2= m^2 + \vec{p}^2$
be replaced by
\begin{equation}
E^2= m^2 + \vec{p}^2 + f(\vec{p}^2,E,m;L_p)
~.
\label{eq:disp}
\end{equation}

If the function $f$ is nonvanishing and nontrivial and
the energy-momentum transformation rules are ordinary (the ordinary
Lorentz transformations) then clearly $f$ cannot have the exact
same structure for all inertial observers. In this case one
would speak of an instance in which Lorentz invariance is broken,
and one could assume that, in spite of the deformation
of the dispersion relation,
the rules for energy-momentum conservation would be undeformed.

If instead $f$ does have the exact
same structure for all inertial observers, then necessarily
the transformations between these observers must be deformed
(they cannot be the ordinary linear Lorentz transformation rules).
In this case one
would speak of an instance in which the Lorentz transformations
are deformed, but there is no preferred frame, the theory
is still fully relativistic~\cite{gacdsr}.
Having deformed the transformation rules between
observers one must also necessarily~\cite{gacdsr} deform
the rules for energy-momentum conservation
(these rules are ``laws of physics, in the Galilei sense,
and must therefore be the same for all inertial observers).

Most work in this area has been devoted to the case in which Lorentz invariance
is actually broken, the possibility that Lorentz invariance might be
deformed was introduced only very recently by this
author~\cite{gacdsr,jurekdsr,michele,dsr3,rossano}.
An example in which all details of the deformed Lorentz symmetry
have been worked out is the one in which one enforces
as an observer-indepedent statement the dispersion relation
\begin{equation}
L_p^{-2}\left(e^{L_p E}
+e^{- L_p E}-2\right)-\vec{p}^2 e^{-L_p E}
=m^2
~.
\label{eq:disptwo}
\end{equation}
In leading (low-energy) order this takes the form
\begin{equation}
E^2 = \vec{p}^2 + m^2 - L_p E \vec{p}^2
~.
\label{eq:displead}
\end{equation}
The Lorentz transformations and the energy-momentum conservation
rules are accordingly modified~\cite{dsr3}.

While the case of deformed Lorentz symmetry might exercise a stronger
conceptual appeal (since it does not rely on a preferred class of inertial
observers), for the purposes of this paper it is sufficient to consider
the technically simpler (and, by the way, still more popular in the
quantum-gravity community)
context of broken Lorentz invariance.
Upon admitting a breakup of Lorentz invariance
it becomes legitimate, for example, to adopt
the dispersion relation (\ref{eq:displead})
without deforming the rules for energy-momentum conservation.
I will use this scenario to illustrate how a tiny (Planck-length suppressed)
effect, such as the one described by
(\ref{eq:displead}), could be observed in certain experimental
contexts.

\subsection{In-vacuo dispersion}
\noindent
A deformation term of order $L_p E^3$ in
the dispersion relation, such as the one in (\ref{eq:displead}),
leads to a small energy dependence of the speed of photons
of order $L_p E$, by applying the relation $v = dE/dp$.

An energy dependence of the speed of photons
of order $L_p E$ is completely negligible in nearly all physical
contexts, but it can be significant~\cite{grbgac,billetal}
in the analysis of short-duration gamma-ray bursts that reach
us from cosmological distances.
For a gamma-ray burst a typical estimate of the time travelled
before reaching our Earth detectors is $10^{17} s$.
Microbursts within a burst can have very short duration,
as short as $10^{-4} s$.
We therefore have one of the ``amplifiers" mentioned in Section~1:
the ratio between time travelled by the signal and time structure
in the signal is a (conventional-physics) dimensionless
quantity of order $\sim 10^{17}/10^{-4} = 10^{21}$.
It turns out that this ``amplifier" is sufficient to study
energy dependence of the speed of photons
of order $L_p E$. In fact, some of the photons in these bursts
have energies in the $10 MeV$ range and higher.
For two photons with energy difference of order $10 MeV$ an $L_p E$
speed difference over a time of travel of $10^{17} s$
leads to a relative time-delay on arrival that is of order $10^{-4} s$,
which would be detected~\cite{grbgac,billetal}
upon comparison of the structure of the signal
in different energy channels.
The next generation of gamma-ray telescopes,
such as GLAST~\cite{glast},
will exploit this idea to search for
energy dependence of the speed of photons
of order $L_p E$.

\subsection{Modified thresholds}
\noindent
Let us now consider another significant prediction
that comes from adopting
the dispersion relation (\ref{eq:displead}).
While in-vacuo dispersion, discussed in the preceding Subsection,
only depends on the deformation of the dispersion relation,
the effects considered in this Subsection (and the next)
also depends
on the rules for energy-momentum conservation, which, as announced,
I shall for simplicity assume to be unmodified.

The point I want to make here is that also certain types of energy
thresholds for particle-production processes may be sensitive
to the tiny $L_p E^3$ modification of
the dispersion relation I am considering for illustrative purposes.

Let us focus on a collision between
a soft photon of fixed/known energy $\epsilon$
and a high-energy photon of energy $E$, whose value is to be determined
assuming the conditions for threshold electron-positron pair production
are met.
It is useful to review briefly the usual calculation
of the $E$ threshold.
One can optimize the calculation by starting with the
observation that the photon-photon invariant
evaluated in the lab frame
should be equal to (among other things) the electron-positron
invariant evaluated in the center-of-mass frame:
\begin{equation}
(E+\epsilon)^2 - (P-p)^2 = 4 m_e^2
~,
\label{throne}
\end{equation}
which, after using the ordinary dispersion relation,
turns into $4 E \epsilon =4 m_e^2$.
So the threshold condition is
\begin{equation}
E \epsilon = m_e^2
~.
\label{thrtwo}
\end{equation}
Notice that in going from (\ref{throne}) to (\ref{thrtwo})
using the ordinary dispersion relation the leading-order
terms of the type $E^2$ have cancelled out, leaving behind the
much smaller (if $\epsilon \ll E$) term of order $E \epsilon$.
This cancellation provides the ``amplifier".
The ``amplifier" is $E/\epsilon$.
If the threshold condition in modified at order $L_p E^3$
the modification will be significant if $L_p E^3$ is
comparable to $E \epsilon$.
While we normally expect $L_p$-related effects to become significant
when the particles involved have energy $1/L_p$, here the effect
is already significant when $E \sim (\epsilon/L_p)^{1/2}$,
which can be considerably smaller than $1/L_p$ if $\epsilon$ is small.
In the specific case
of the deformed dispersion relation (\ref{eq:displead}),
applying ordinary energy-momentum conservation
one finds~\cite{gactp} the modified threshold relation
\begin{equation}
E \epsilon - L_p {E^3 \over 8}= m_e^2
~.
\label{thrTRE}
\end{equation}
For $E \sim 10 TeV$ and $\epsilon \sim 0.01 eV$
the modification of the threshold is already significant.
These values of $E$ and $\epsilon$ are relevant for the observation
of multi-$TeV$ photons from certain Markarians~\cite{aus,gactp}.
This high-energy photons travel to us from very far
and they travel in an environment populated by soft photons,
some with energies suitable for acting as targets
for the disappearance of
the hard photon into an electron-positron pair.
Depending on some properties (such as the density) of the far-infrared
soft-photon background (which are still not fully known)
the observation
of multi-$TeV$ photons from certain Markarians
may appear to be surprising~\cite{aus}
within conventional relativistic astrophysics.
The Planck-scale induced deformation term
in Eq.~(\ref{thrTRE}), by shifting up the value
of the threshold energy, could explain these observations
from Markarians.

A similar argument can be applied to cosmic rays.
The puzzling fact that cosmic rays are seen above
the GZK limit can also be interpreted as a violation of a
relativistic threshold and again
the dispersion relation (\ref{eq:displead})
combined with conventional energy-momentum conservation
would lead to a prediction for the relevant threshold
(the photopion-production threshold)
in agreement with data~\cite{kifu,gactp}.

These observations, preliminary as they are,
may well be the first ever manifestation of Planck-scale physics.
The fact that we can finally at least contemplate
this hypothesis has
increased interest\footnote{It is not uncommon that preliminary data
generate interest in related theory subjects, and in some cases
the lessons learned through those theoretical studies outlast the possible
negative evolution of the experimental situation. This author is
familiar~\cite{bjpap} with the theory work that was motivated
by the so-called ``centauro events". It is now widely believed that
centauro events were a ``mirage", but in the process we did learn that
the formal structure of QCD allows the vacuum to be temporarily
misaligned (disoriented chiral condensates) and the RHIC collider
is conducting dedicated experiments.}$~$
in the whole quantum-gravity phenomenology
research programme.

\subsection{Modified decay amplitudes}
\noindent
But this is not all. There is even a third opportunity for
doable experiments to look for a manifestation of
the tiny, Planck-length suppressed,
modification of the dispersion relation which I am considering
as illustrative example of quantum-gravity phenomenology
exercise. This third opportunity has to do
with particle-decay amplitudes and I shall discuss it through
the example of the decay of a pion into two photons.
First let us try to understand why pion-decay into two photons
could be so sensitive. This is interesting because in this
case the ``amplifier" takes an unexpected form.

Again it is useful to review the relevant derivation
within ordinary relativistic kinematics.
One can optimize the calculation by starting with the
observation that the photon-photon invariant in the lab frame
should be equal to the pion
invariant:
\begin{equation}
(E+E')^2 - (\vec{p}+\vec{p}')^2 = m_\pi^2
~.
\label{pionone}
\end{equation}
Using the conventional relativistic
dispersion relation this can be easily turned
into a relation between
the energy $E_\pi$ of the incoming pion, the opening angle $\phi$
between the
outgoing photons, and the energy
$E$ of one of the photons
(the energy $E'$ of the second photon
is of course not independent; it is given by
the difference between the energy
of the pion and the energy of the first photon):
\begin{eqnarray}
\cos(\phi) &\! = \!& {2 E E' - m_\pi^2
\over
2 E E'} ~,
\label{pithreshone}
\end{eqnarray}
where indeed $E' = E_\pi - E$.
The reader should notice that $\cos(\phi) \leq 1$,
as required by the fact that $\phi$ is a real physical angle,
for all values of $E$.
Note however that typically (unless $E \simeq 0$ or $E \simeq E_\pi$)
$m_\pi^2 \ll 2 E E' \sim E_\pi^2/2$
and the equation for $\cos(\phi)$
as the form $\cos(\phi) = (2 E E' - \Delta)/2 E E'$.
So the fact that $\cos(\phi) \leq 1$ for all values of $E$
depends only on the fact that $\Delta > 0$, which is authomatically
satisfied within ordinary relativistic kinematics through the
prediction $\Delta = m_\pi^2$. A new kinematics predicting
that $\Delta < 0$ for some values of $E$ would have
significant implications, and in order to render $\Delta$ negative
it is sufficient to introduce a relatively small correction,
a correction of order $m_\pi^2$.

This is what happens in the scheme I am considering.
The deformed dispersion relation (\ref{eq:displead}),
when combined with ordinary energy-momentum conservation,
modifies the relation between $\phi$, $E_\pi$ and $E$
according to the formula~\cite{gacpion}
\begin{eqnarray}
\cos(\phi) &\! = \!& {2 E E' - m_\pi^2
+ 3 L_p E_\pi E E'
\over
2 E E' + L_p E_\pi E E'} ~.
\label{pithresh}
\end{eqnarray}
This relation shows that at high energies the phase space available to the decay
is anomalously reduced:
for given value of $E_\pi$ certain values of $E$
that would normally be accessible to the decay are no longer
accessible (they would require $cos \theta > 1$).
This anomaly starts to be noticeable at pion energies of
order $(m_\pi^2/L_p)^{1/3} \sim 10^{15}eV$, but only
very gradually (at first only a small portion of the available
phase space is excluded).
Remarkably, this type of behaviour could explain~\cite{gacpion} certain
puzzling features~\cite{dedenko}
of the longitudinal development of the air showers produced by certain
high-energy cosmic-rays.

Independently of whether or not this preliminary experimental
encouragement is confirmed by more refined data on pion decay,
it is important for the line of argument presented in this paper
that this scheme for the analysis of pion decay is another
example of a Planck-scale scheme in which the effects
become significant well below the Planck scale.
The effects are already significant at pion energies of
order $(m_\pi^2/L_p)^{1/3} \sim 10^{15}eV$.
The careful reader will notice that in this case the ``amplifier"
is $E_\pi/m_\pi$.

\section{Laser-interferometric Foam Studies:
an example of Studies of nonsystematic Quantum-Gravity Effects}
\noindent
In this Section I illustrate, focusing on effects associated
with distance fuzziness, the type of issues that emerge
in the analysis of nonsystematic quantum-gravity
effects.
Some differences with respect to the method of analysis
of systematic effects will emerge.

\subsection{Distance fuzziness}
\noindent
Let us consider the possibility that the concept of distance
be fuzzy in the intuitive sense of spacetime foam studies
(and in the technical operative sense of Section~2).
A robust analysis would require some model of
this fuzziness and of the mechanisms that bring it about,
but top-to-bottom theories provide very little guidance
on this point. The type of systematic effects
analyzed in the previous Section is governed by symmetry principles,
and on those at least some preliminary (and vague) guidance
can be gotten from top-to-bottom theories,
but on distance fuzziness
we lack even that level of guidance.

I resort here to a strictly phenomenological approach.
Let us consider an experiment in which a distance $L$ plays
a key role, meaning that one is either measuring $L$ itself
or the observable quantity under study depends strongly on $L$.
Let us assume (as commonly done)
that in quantum gravity there should be a fundamental limitation on
the measurability of $L$, a new uncertainty principle, and
let us characterize this limitation in terms
of a mean square deviation $\sigma_L^2$.
We will want to analyze the implications of various hypotheses
for $\sigma_L^2$ and compare them to the type of sensitivities that
are achievable in relevant experiments.

Two experimental contexts which could be
promising in this respect are: the gamma-ray-burst
context already considered in the previous Section, where a very large
distance is involved but the spread of times of arrival is relatively
small, and the context of laser interferometry, where a relatively
large distance can be monitored with extreme accuracy.

If $\sigma_L^2$ is independent of the time of observation
(and therefore independent of $L$) one is naturally
led to the estimate $\sigma_L^2 \sim L_p^2$.
It is easy to verify that this estimate of $\sigma_L^2 \sim L_p^2$
would not be observably large, even in our two most promising
experimental contexts (the relevant sensitivities are several orders
of magnitude below the required level).

If one goes beyond the constant-$\sigma_L^2$ assumption, it is natural
to consider also a possible
dependence of $\sigma_L^2$ on the time $T$ of observation of $L$ required
by the experiment. This can be motivated in various ways~\cite{gacgwi},
and it is in the spirit of certain discretized
mechanisms of space-geometry
time evolution that are emerging within the loop-quantum-gravity
research programme (see, {\it e.g.}, Refs.~\cite{fotini,ambjorn}).
Introducing dimensionless parameters $A$, $B$
(to be determined experimentally) one can then write $\sigma_L^2$ as
\begin{equation}
\sigma_L^2 \simeq A L_p^2 + B L_p c T
~.
\label{sigmalaser}
\end{equation}
The previous remark on the case in which $\sigma_L^2$
is $T$-independent
means that experimental limits on $A$ are
not significant.
Let us consider the limits on $B$,
and let us start with the context of gamma-ray bursts.
As mentioned,
for a gamma-ray burst a typical estimate of the time travelled
before reaching our Earth detectors is $10^{17} s$
and microbursts within a burst can have very short duration,
as short as $10^{-4} s$.
It is easy to realize that this imposes that
whatever fundamental ``uncertainty" affects the relevant
distance $c {\cdot} 10^{17} s$ it cannot be bigger than $c {\cdot} 10^{-4} s$.
This corresponds to a limit on $B$ which is of order $B < 10^{13}$.
This limit does not appear to be particularly interesting,
but in the other context, the one of laser interferometry,
a somewhat more encouraging estimate emerges.

\subsection{Laser-interferometric limits}
\noindent
For the context of gamma-ray bursts the ``amplifier" of distance fuzzines
is of order $10^{21} = (10^{17} s)/(10^{-4} s)$.
A superficial analysis of modern laser interferometers
would attribute to them a comparable ``amplifier" estimate.
In fact, one major and well-known quality of these modern
interferometers
(whose primary objective is the discovery of the
classical-physics phenomenon of gravity waves)
is their ability to detect gravity waves
of amplitude $\sim 3 {\cdot} 10^{-19}m$
by careful monitoring of distances of order $\sim 3 {\cdot} 10^3 m$.
This would lead
to an ``amplifier" which is of order $10^{22}$.
However, the correct way to characterize the sensitivity
of an interferometer requires~\cite{gacgwi,bignapap,saulson}
the analysis of
the power spectrum of the strain noise which is left over after
all the sophisticated noise-reduction techniques have been applied.
In modern interferometers this strain power-noise spectrum
is of order $10^{-44} Hz^{-1}$ at observation frequencies of
about $100 Hz$,
and in turn this implies~\cite{gacgwi} that for a gravity wave
with $100 Hz$ frequency the detection threshold is indeed around
$\sim 3 {\cdot} 10^{-18}m$.
But not all fluctuation mechanisms are smooth waves.
An ideal wave deposits all its energy in the frequency
band of observation that includes its own frequency of oscillation.
Things work differently for other fluctuation mechanisms,
and particularly for discrete fluctuation mechanisms.

The ansatz $\sigma_L^2 \sim B L_p T$, on which I am focusing for
illustrative purposes, has the time dependence characteristic
of random-walk process. Indeed one obtains $\sigma_L^2 = L_p T$
by assuming
that the
distances $L$ betweeen the test masses of an interferometer be
affected by Planck-length
fluctuations of random-walk type
occurring at a rate of one per Planck time ($\sim 10^{-44} s$).
It is easy to verify~\cite{gacgwi}
that such fluctuations would induce strain noise
with power spectrum given by $L_p L^{-2} f^{-2}$.
For $f \sim 100 Hz$ and $L\sim 3 {\cdot} 10^3 m$ this corresponds
to strain noise at the level $10^{-37} H\!z^{-1}$,
well within the reach of the sensitivity of
modern interferometers.

Fluctuations genuinely at the Planck scale
(the simple scheme I used to illustrate my point
involves Planck-length fluctuations occurring
at a rate of one per Planck time)
can lead to an effect that, while being very small in absolute
terms, is large enough for testing with modern interferometers.
The careful reader will realize that this is due to
the fact that a meaningful estimate of the ``amplifier"
in laser interferometers is obtained by combining
the characteristic frequency of observation
and the noise levels aspected within
conventional physics $1/( f {\cdot} 10^{-44} Hz^{-1}) \sim 10^{42}$.

\subsection{Significance of the laser-interferometric limits}
\noindent
When discussing this type of experimental programmes
at conferences and similar occasions, one is often invited to
express an opinion on the significance of the
forthcoming laser interferometers for spacetime-foam studies.
Of course, such an opinion is beyond the scopes of
quantum-gravity phenomenology.
This type of exercise in quantum-gravity phenomenology can only
identify large ``amplifiers" and establish their possible connection
with effects that are of Planck-length magnitude.
The next step\footnote{Actually, there could be an intermediate step
between the strictly phenomenological dimensional analysis (without
even a naive picture of the supporting mechanism) and the
analysis within a ``promising" theory: one could construct a naive picture
of a quantum-gravity mechanism generating laser-intereferometric noise.
But even this intermediate step is not easily taken in this context,
because of the lack of any guidance, even just in terms of
some symmetry principle. For example, one can analyze quantum-gravity
noise in terms of noise in the paths of each of the $N_\gamma$
individual photons
of the beam, but this would not be a representation of a fundamental
noise (not in the sense of the
noise that ordinary quantum mechanics contributes
to laser interferometry) because it would inevitably disappear in
the $N_\gamma \rightarrow \infty$ limit (whereas the contribution
to laser-interferometric noise
of ordinary quantum mechanics is composed of two terms~\cite{saulson},
one descreasing with $N_\gamma$ and one growing with $N_\gamma$).
Still, as I shall discuss in detail
in Ref.~\cite{bignapatwo}, this picture in terms of the paths of
individual photons might encourage interferometric studies
with low-power lasers, in order to reduce the apparent suppression
of the effect by $\sqrt{N_\gamma}$.}$~$
would be to analyze the physical context
in terms of a ``promising quantum-gravity theory".
This step cannot be taken for not one, but two reasons:
(i) we have no quantum-gravity theory whose ``promise"
relies on the successfull prediction of some experimentally
verified experimental facts, and (ii) even if we wanted to
attribute ``promise" to the theories which have developed
into appealing conceptual/mathematical structures,
such as loop quantum gravity and string theory,
we are faced with the fact that these theories are
still unprepared to provide this type of physical estimates.

One interesting way to address the issue of ``significance"
can be based on attempting to address the following question:
is the next generation of laser interferometers really entering
a new region of exploration? (are the new limits to be obtained
in those experiments crossing some meaningful sensitivity boundaries?)
In this sense one can state that these forthcoming experiments
are significant. This is best stated by writing a phenomenological
formula for the strain noise power spectrum:
\begin{equation}
\rho_h(f) = {\alpha L_p \over c} + {L_p \over \Lambda_1 f}
+ {c L_p \over \Lambda_2^2 f^2}  + {} +~...
~,
\label{rholaser}
\end{equation}
where $\alpha, \Lambda_1,\Lambda_2$ parametrize our ignorance of the
coefficients (they should be predicted by theory or measured)
and the choice of notation emphasizes the fact that the first term
requires a dimensionless coefficient, while the second term and
the  third
term require coefficients
with units of inverse-length and inverse-square-length respectively.

The three terms included in (\ref{rholaser}) are just indicative.
A fully general phenomenolgoical formula should involve
many more types of $f$ depedence and the possibility that the
spectrum might not be linear in $L_p$ ({\it e.g.} it could
go like $L_p^2$).
However, (\ref{rholaser}) allows us to characterize in phenomenological
quantitative terms the type of sensitivity thresholds that
are being reached with the next generation of laser interferomters.
The next generation of laser interferometers
will have sensitivity that goes down to $\alpha$
even smaller than 1, whereas until, say, a decade ago the sensitivity
was several orders of magnitude away from $\alpha =1$.
Similarly the next generation will have sensitivity that goes down
to values of $\Lambda_1$
and $\Lambda_2$ as large as the optical length of the arms of
the interferometer, whereas until a decade ago the sensitivity
was several orders of magnitude away from these levels.

The lack of theoretical guidance does not allow us
to form any justifiable opinion about the ``theory significance"
of the limits that will be obtained by the next generation of
laser interferometers, but at the phenomenological level
we can recognize that some meaningful sensitivity thresholds
will be reached by these forthcoming experiments
(and instead these sensitivity thresholds were totally unaccessible
to the previous generation of laser interferometers).
Perhaps most important for future developments is the fact
that this analysis exposed the fact that a meaningful estimate
of the ``amplifier" in laser interferometers must depend strongly
on whether the fluctuations are smooth or discretized.
For discretized fluctuation mechanisms the ``amplifier"
could be as high as $10^{42}$.
This is particularly significant since one of the objectives of
quantum-gravity phenomenology (see Section~1) should be the one
of establishing which experimental contexts are able to set
the most stringent limits (independently of the ``significance"
of these limits, which requires at this stage some opinion
about the workings of quantum gravity) on each of the
effects that have surfaced in the quantum-gravity literature.
The list of such effects is very limited
(only very few entries) and it seems necessary from a
strictly scientific viewpoint, to establish where we
are in the exploration of each of these effects.
The observations reported here
(and in Refs.~\cite{gacgwi,bignapap,bignapatwo})
are in this sense noteworthy since they indicate that in some
theories of quantum gravity laser-interferometric limits
on distance fuzziness will be more stringent than the
corresponding limits obtainable with gamma-ray-burst analysis,
in spite of the fact that a naive estimate of the
amplifiers in these two contexts would suggest that
they have comparable sensitivity.

\subsection{Futility of the Salecker-Wigner debate}
\noindent
The phenomenology of distance fuzziness is, as emphasized,
already rendered more delicate by the fact that top-to-bottom
theories are totally unable to provide us any guidance
(whereas in the case of the systematic effects considered
in the previous Section one could at least rely on some
emerging intuition concerning the faith of
Lorentz symmetry).
Somehow the debate got also partly penalized by the attention
some authors devoted to a potentially relevant argument
due to Salecker and Wigner~\cite{sw},
which is however not needed in order to justify a
quantum-gravity analysis of laser interferometers.

The Salecker-Wigner argument is a heuristic argument
which leads to some intuition
on new limits on the measurability of distances,
something which of course is potentially relevant
for the issue here being considered.
In previous papers on this subject this author did
mention this Salecker-Wigner argument as one
of the arguments motivating interest in these laser-interferometric
studies. The Salecker-Wigner argument, as all heuristic arguments,
is not immune from criticism or at least skepticism,
and it is of course not surprising that in Refs.~\cite{adlerSW,baezSW}
certain alternative ideas (actually rather naive ideas~\cite{gacSW},
but this is not the point here)
on how the Salecker-Wigner setup should be properly analyzed
were presented. What is really surprising is that, motivated exclusively
by these views on the Salecker-Wigner setup, Refs.~\cite{adlerSW,baezSW}
argued that phenomenological interest in
laser-interfeometric studies would not be justifiable.
As shown above a certain level of interest (although, as emphasized,
not necessarily in the sense of ``theory significance")
in these laser-interferometric studies can be justified at a strictly
phenomenological level, without any reference to the Salecker-Wigner
argument. This point had already been articulated in detail
in Ref.~\cite{bignapap} (which preceded Refs.~\cite{adlerSW,baezSW})
but was somehow missed by the authors of Refs.~\cite{adlerSW,baezSW}
(which in fact do not include a citation of Ref.~\cite{bignapap}
in their reference lists).

As a way to reduce the confusion generated by the peculiar
development of this debate involving the Salecker-Wigner argument and
laser-interferometric studies,
I have here chosen to refrain completely from any reference
to the Salecker-wigner argument in motivating
laser-interferometric quantum-gravity phenomenology.

\section{Outlook}
\noindent
A good measure of the pace of development of quantum-gravity
phenomenology can be obtained by comparing the number
of ideas that needed to be covered by this author in the
previous review some three years ago~\cite{polonpap}
and the number of ideas that have been covered
(however briefly) in this review.
Moreover, even some of the ideas that were already
under consideration three years ago are now studied
and understood at a much deeper level.

On the theory side, perhaps the most promising research
programme remains the one concerning symmetries.
It appears safe to bet that over the next few years
the theoretical study of the faith of classical-spacetime symmetries
in quantized (discretized, noncommutative,...)
descriptions of spacetime will produce even more
exciting results.

But for the next ten years we should be even more
optimistic about progress
on the experimental front.
Laser interferometry, discussed in the previous Section,
will go through a remarkable upgrade through
the operation of LIGO, VIRGO, and (hopefully) LISA.
Although at a slower pace also tests of CPT symmetry
will keep improving.

Perhaps the most exciting data we will get are the ones
pertaining Lorentz symmetry.
The preliminary evidence of deviations from
conventional Lorentz invariance (here discussed in Section~3),
which resides primarily (and most robustly)
in cosmic-ray observations, will be put under severe scrutiny
with new cosmic-ray observatories, such as Auger.
If the preliminary evidence is confirmed by these more powerful
cosmic-ray observatories one could then support (or disprove)
the new-kinematics interpretation of the GZK puzzle through
the results of searches of the corresponding in-vacuo dispersion
with the next generation of gamma-ray observatories,
such as GLAST~\cite{glast}.

\nonumsection{Acknowledgments}
\noindent
I am particularly greatful to Dharam V.~Ahluwalia and Naresh Dadhich
for their role in organizing the IGQR-I meeting.
The focus of that meeting on observable aspects of quantum gravity
was an important contribution to the development of the field.
It was truly disappointing for me having to miss the meeting
at the last minute (and Dharam was kind enough to make me
regret it even more with his descriptions of the
pleasant atmosphere
and scientific intensity of the meeting).
I also thank the Perimeter Institute for Theoretical Physics
for hospitality during part of my work on this manuscript.

\nonumsection{References}

\end{document}